\documentclass[runningheads]{llncs}
\usepackage{graphicx}
\usepackage[dvipsnames]{xcolor}
\usepackage{float}
\usepackage{subfigure}
\usepackage{cite}
\usepackage{multirow}

\newif\ifdraft
\draftfalse 

\ifdraft
\def\H#1{\textcolor{Orange}{#1}}
\def\Hc#1{\textcolor{Orange}{\textit{\textsf{ \small [HB: #1]}}}}  
\def\Hd#1{\textcolor{purple}{\textit{[deleted: #1]}}}  

\else
\def\H#1{#1}
\def\Hc#1{}  
\def\Hd#1{}  
\fi

\usepackage{array}
\newcolumntype{L}[1]{>{\raggedright\let\newline\\\arraybackslash\hspace{0pt}}m{#1}}
\newcolumntype{C}[1]{>{\centering\let\newline\\\arraybackslash\hspace{0pt}}m{#1}}
\newcolumntype{R}[1]{>{\raggedleft\let\newline\\\arraybackslash\hspace{0pt}}m{#1}}

\begin{document}
%
\title{Multiclass segmentation as multitask learning for drusen segmentation in retinal optical~coherence~tomography}

\titlerunning{Multitask learning for drusen segmentation in OCT scans}
%
\author{Rhona Asgari 
\and Jos\'e Ignacio Orlando
\and Sebastian~Waldstein
\and Ferdinand~Schlanitz
\and Magdalena~Baratsits
\and Ursula~Schmidt-Erfurth
\and Hrvoje~Bogunovi\'c
}

\authorrunning{Asgari et al.}
%
\institute{Christian Doppler Laboratory for Ophthalmic Image Analysis, Department of Ophthalmology, Medical University of Vienna,
Vienna, Austria}

%
\maketitle              

\begin{abstract}
    Automated drusen segmentation in retinal optical coherence tomography (OCT) scans is relevant for understanding age-related macular degeneration (AMD) risk and progression.
    This task is usually performed by segmenting the top/bottom anatomical interfaces that define drusen, the outer boundary of the retinal pigment epithelium (OBRPE) and the Bruch's membrane (BM), respectively. 
    In this paper we propose a novel multi-decoder architecture that tackles drusen segmentation as a multitask problem. Instead of training a multiclass model for OBRPE/BM segmentation, we use one decoder per target class and an extra one aiming for the area between the layers. We also introduce connections between each class-specific branch and the additional decoder to increase the regularization effect of this surrogate task. We validated our approach on private/public data sets with 166 early/intermediate AMD Spectralis, and 200 AMD and control Bioptigen OCT volumes, respectively. Our method consistently outperformed several baselines in both layer and drusen segmentation evaluations.
\end{abstract}

\section{Introduction}
\label{sec:introduction}

Age-related macular degeneration (AMD) is one of the leading causes of blindness in elderly population in the developed world~\cite{Wong2014}.
One of the first clinical hallmarks of AMD is the presence of drusen, waste material accumulations in the area delimited by the outer boundary of the retinal pigment epithelium (OBRPE) and the Bruch's membrane (BM). 
Optical coherence tomography (OCT) is the state-of-the-art imaging modality to assess AMD patients, as it allows to visualize the retinal layers and study pathological changes due to AMD, including drusen. Segmenting drusen in OCT is relevant for quantifying disease progression~\cite{Schlanitz2017}, although doing it manually is costly, tedious and time consuming. Current methods for automated drusen segmentation are based on identifying the OBRPE and BM interfaces, as every non-overlapped area in between those surfaces is considered drusen~\cite{GorgiZadeh2017,fang2017automatic}. Deep learning techniques based on convolutional neural network (CNNs) have been recently explored for this task~\cite{fang2017automatic,GorgiZadeh2017,shah2018multiple,liefers2019dense}. In~\cite{fang2017automatic}, a patch-based CNN is applied for feature extraction, combined with graph search strategies and standard classifiers. In~\cite{shah2018multiple}, an image-level classification CNN is applied to predict the vertical coordinates of each surface. A similar idea is followed in~\cite{liefers2019dense} to predict surface positions using a 2D-to-1D approach.


\begin{figure}[t!]
    \centering
    \includegraphics[width=0.99\textwidth]{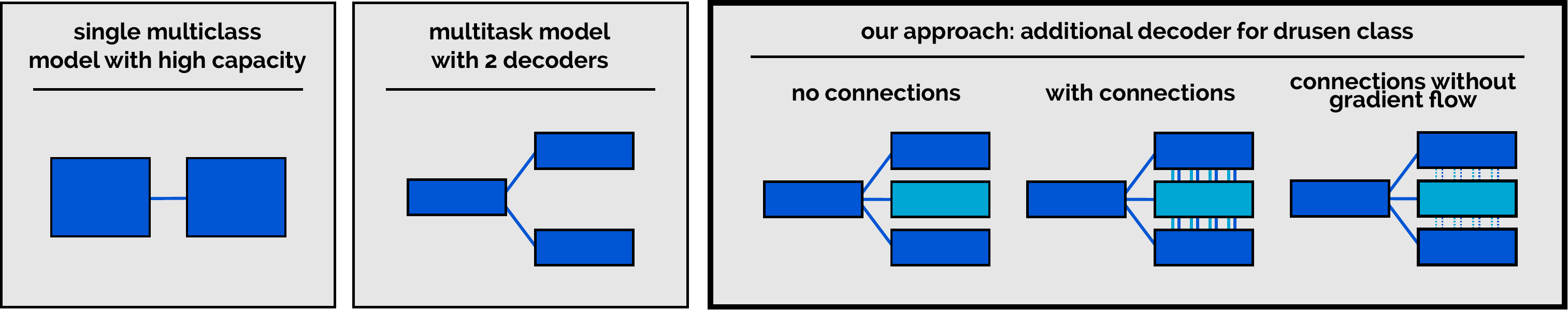}
    \caption{Different multiclass segmentation approaches. From left to right: single multiclass model with high capacity, multitask model with 2 decoders~\cite{playout2018multitask} and our approaches with an additional decoder for drusen class: no connections and surrogate decoder, with connections and gradient flow, and with connections but no gradient flow. Box sizes indicate the capacity of the module.}
    \label{fig:schematic-multiclass}
\end{figure}

Segmenting the OBRPE and the BM is a multiclass segmentation problem that can be tackled in different ways (Fig.~\ref{fig:schematic-multiclass}). 
The common solution~\cite{GorgiZadeh2017} is to use a multiclass model with high capacity (e.g. encoder/decoder architectures such as the U-Net~\cite{Ronneberger2015}) and a multiclass loss function (e.g. cross-entropy). This model learns how to discriminate between classes and which features are needed to identify them. This interaction cannot be explicitly controlled, and it is not possible to assign portions of capacity to specific classes (although it can be approached by weighting classes in the loss function, e.g. under class imbalance).


An alternative is to pose the multiclass segmentation task as a multitask learning problem. Instead of having a single encoder and a single decoder, both shared among the $K$ target classes, the architecture is split into $K$ decoders. Each decoder is focused on a binary segmentation task, providing class-specific capacities that are exclusively dedicated to the target problem. Simultaneously, a single encoder is shared among them to characterize common features. This architecture can be trained using a linear combination of $K$ binary loss functions, and benefits from the inductive bias of each task by updating the encoder parameters. This approach has been previously explored in~\cite{playout2018multitask} for red/bright lesion segmentation in fundus images, with promising results.

In this paper we bring the multitask approach in~\cite{playout2018multitask} one step further. In particular, we exploit the fact that our drusen segmentation task comprises the segmentation of the OBRPE and the BM to capture the area between them. Hence, instead of having one decoder for each layer class, we introduce a third one aiming to segment the region between both layers (Fig.~\ref{fig:architecture}). Our assumption is that this additional branch will aid the encoder to characterize the appearance of both drusen and non-pathological regions where OBRPE and BM are overlapped. We also explore the influence resulting from introducing additional connections between each of the layer decoders and this intermediate one, with and without gradient flow. Allowing gradient flow is expected to further transfer the inductive bias of the intermediate task not only to the encoder path but also through the main decoders. On the other hand, gradient blocking allows each task to exploit only the feature maps learned for their respective target class.

We experimentally validated our three approaches using private and public data sets of retinal OCT scans with $166$ and $200$ volumes acquired using Spectralis and Bioptigen devices, respectively. Our proposed architectures reported the best performance for drusen segmentation compared to several baselines.



\section{Methods}
\label{sec:methods}

\subsection{Multiclass segmentation as multitask learning}
\label{subsec:our-method}

Given an input image $\textbf{x} \in \mathcal{X}$, our goal is to produce a label $y$ for each pixel $x$, in the label space $\mathcal{L} = \{0, 1, ..., K\}$, with $0$ being background and $K$ the total number of classes of interest. This task is usually performed using a supervised deep learning model, $f_\theta(\mathcal{X}) \rightarrow \mathcal{Y}$, where $\mathcal{X}$ and $\mathcal{Y}$ are the set of B-scans and labelings, respectively. The parameters $\theta$ are learned using a training set $S = \{ (\mathbf{x}^{(i)}, \mathbf{y}^{(i)}), 0 < i < N\}$, with $N$ the total number of pairs $(\mathbf{x}^{(i)}, \mathbf{y}^{(i)})$ of training images and their labels, respectively. This is done by minimizing a loss function $J(\mathbf{y},\mathbf{\hat{y}})$, where $\mathbf{y}$ and $\mathbf{\hat{y}}$ are the manual and predicted multiclass segmentations.

Network parameters can be decomposed as the union of the weights of the encoder $\theta_{E}$ and the decoder $\theta_{D} $, $\theta = \theta_{E} \cup \theta_{D}$. In a typical multiclass setting, all the parameters are shared among classes and it is not possible to assign part of them to each class. In~\cite{playout2018multitask}, the authors proposed to replace the unique decoder by two decoders, one per target. Formally, this is equivalent to model $\theta_{D} = \bigcup_{k=1}^K {\theta}^{k}_{D}$, where each ${\theta}^{k}_{D}$ is the set of parameters of the decoder for the $k$-th class. This model is trained by means of a multitask loss function, which is defined as a linear combination of binary segmentation losses $J(\mathbf{y},\mathbf{\hat{y}}) = \sum_{k=1}^K \lambda_k J_k(\mathbf{y}_k,\mathbf{\hat{y}}_k)$. $\lambda_k$ denotes a weight for the $k$-th loss function, while $\mathbf{y}_k$ and $\mathbf{\hat{y}}_k$ are binary predictions for each class $k$ vs. every other, including the background class.

\subsection{Drusen segmentation in OCT as multitask learning}
\label{subsec:drusen-segmentation}

A retinal OCT scan is a 3D volume composed of consecutive 2D images or B-scans, captured by means of low-coherence interferometry.
We seek the model $f$ to produce a labeling for each input B-scan $\mathbf{x}$. The classical strategy to do so is to aim for the OBRPE and the BM interfaces: a multiclass segmentation problem with background vs. $K=2$ classes~\cite{GorgiZadeh2017,shah2018multiple}. In healthy cases, both classes are overlapped and there is no region in between them. In early/intermediate AMD cases, however, a third class $k=3$, is implicit in the non-overlapped areas of these layers. Instead of considering this class as background, we propose to learn its properties by considering $K=3$ during training. This is done by: (i) adding an extra decoder for this class, and (ii) incorporating another binary segmentation term to the loss function that penalizes errors in the segmentation of this region. Our hypothesis is that the existing decoders will benefit from the inductive bias of this new task through the gradient updates in the encoder, better characterizing the non-overlapped cases. To further increase the regularization effect of this extra task, we also propose to incorporate inbound/outbound connections between each ${\theta}^{k}_{D}$ and ${\theta}^{3}_{D}$, with $k \neq 3$, both with and without gradient~flow. 

\begin{figure}[t!]
    \centering
    \includegraphics[width=\textwidth]{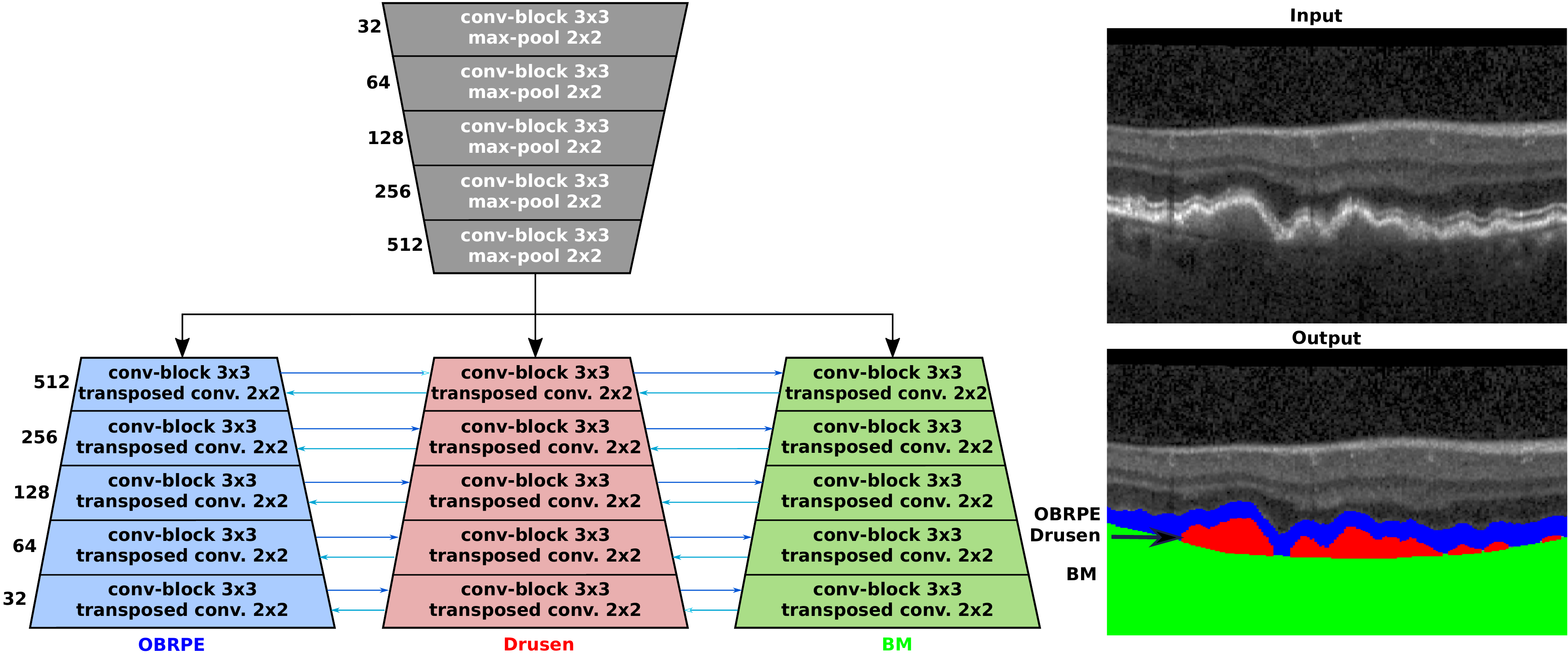}
    \caption{Our multitask segmentation network for layer/drusen segmentation in OCT. Skip connections between each parallel encoder/decoder block were omitted for clarity. The number of output filters of each block are denoted on their left side.}
    \label{fig:architecture}
\end{figure}

Fig.~\ref{fig:architecture} depicts our U-shaped architecture. Skip connections between each parallel encoder/decoder convolutional/deconvolutional blocks were omitted for clarity. Each convolutional block comprises two convolutional layers with $3\times3$ pixels convolutions. The Generalized Dice loss function was applied for each $J_k(\mathbf{y}_k,\mathbf{\hat{y}}_k)$~\cite{Crum2006}, and the predicted labels were binary masks for each target class, as depicted in Fig.~\ref{fig:architecture}. These outputs were combined to produce the final segmentation. Any segmented component with an area smaller than expected size for BM region (yellow in Fig.~\ref{fig:architecture}) or for RPE region (red in Fig.~\ref{fig:architecture}), in the output mask was treated as noise and removed from the mask. To retrieve the surfaces of the BM/OBRPE layers, a postprocessing strategy was applied: for each vertical column in the B-scan (or A-scan), the first/last row of activated pixels was taken as the surface boundary, respectively. 

\section{Experimental setup}
\label{sec:materials-and-evaluation-metrics}




Our private data set consisted of 560 fovea-centered Spectralis OCT volumes of patients with early/intermediate AMD. In total, there are $51K$ B-scans acquired from 48 patients and 63 eyes. Each scan comprises $1024 \times 97 \times 496$ voxels, with a voxel size of $5.7\times 60.5 \times 3.87 \mu m^3 $, covering the field of view of $6\times6\times2$ mm$^3$. A manual labelling was produced for each B-scan on every volume. The Iowa Reference Algorithms~\cite{chen2012three} were used to generate a first raw segmentation, which was subsequently manually corrected by an expert clinician. To train and evaluate the networks, we split the data into 41K B-scans for training and validation (33 patients, 42 eyes) and 10K B-scans for testing (15 patients, 21 eyes). Scans from the same subjects were always placed in the same subset.

A second evaluation was performed on a publicly available data set from Duke~\cite{farsiu2014quantitative}, comprising 384 OCT volumes from 269 patients with intermediate AMD and 115 control subjects. Images were acquired with a Bioptigen device, where each scan consists of $1000 \times 100 \times 512$ voxels, with a voxel size of $6.54\times 67 \times 3.23 \mu m^3 $, covering the field of view of $6.7\times6.7\times1.66$ mm$^3$. The labels in this data set comprise both the BM and the inner boundary of the RPE (IBRPE). In healthy cases, the region between these two layers is not empty but covers the RPE, while in diseased cases it accounts for the RPE+drusen complex. The data is divided into 165 AMD plus 5 Control,  
In order to be consistent with the recent literature, we split the data set into training, validation and test sets using the same proportions as recently used in~\cite{liefers2019dense}. Notice, however, that this does not ensure that the same images are used for comparison. We performed the evaluation both in control and AMD diseased cases.


\paragraph{Training Setup} Our method and the baselines were trained with a batch size of 16 for at least 50 epochs, using Adam optimization with an initial learning rate of $\eta = 10^{-5}$, decreased by a factor of $10^{-7}$ after every epoch. Training was halted if no improvement in loss function(dice coefficient) was observed in 4 consecutive epochs. Each input B-scan was resized to 256$\times$256 pixels and normalized to zero mean and unit variance. Data augmentation was applied in the form of flipping and translation. An equal weighting $\lambda_k = 1$ was used for each loss function $J_k$. 


\section{Results}
\label{sec:results}

We evaluated our method in terms of drusen and layer segmentation performance. For drusen segmentation, we used classical binary evaluation metrics such as Dice index, precision and recall (sensitivity). For layer segmentation, we used mean absolute error. Since our data set includes multiple scans for the same eye of different patients, the evaluation metrics were first computed at an eye level and then averaged by the number of visits. This ensures to have independent samples for subsequent statistical analysis. The significance of the results was evaluated using a paired Wilcoxon signed-rank test with $\alpha=0.05$. 

\begin{figure}[t]
  \centering
  \subfigure[OBRPE]{\includegraphics[width=0.47\textwidth]{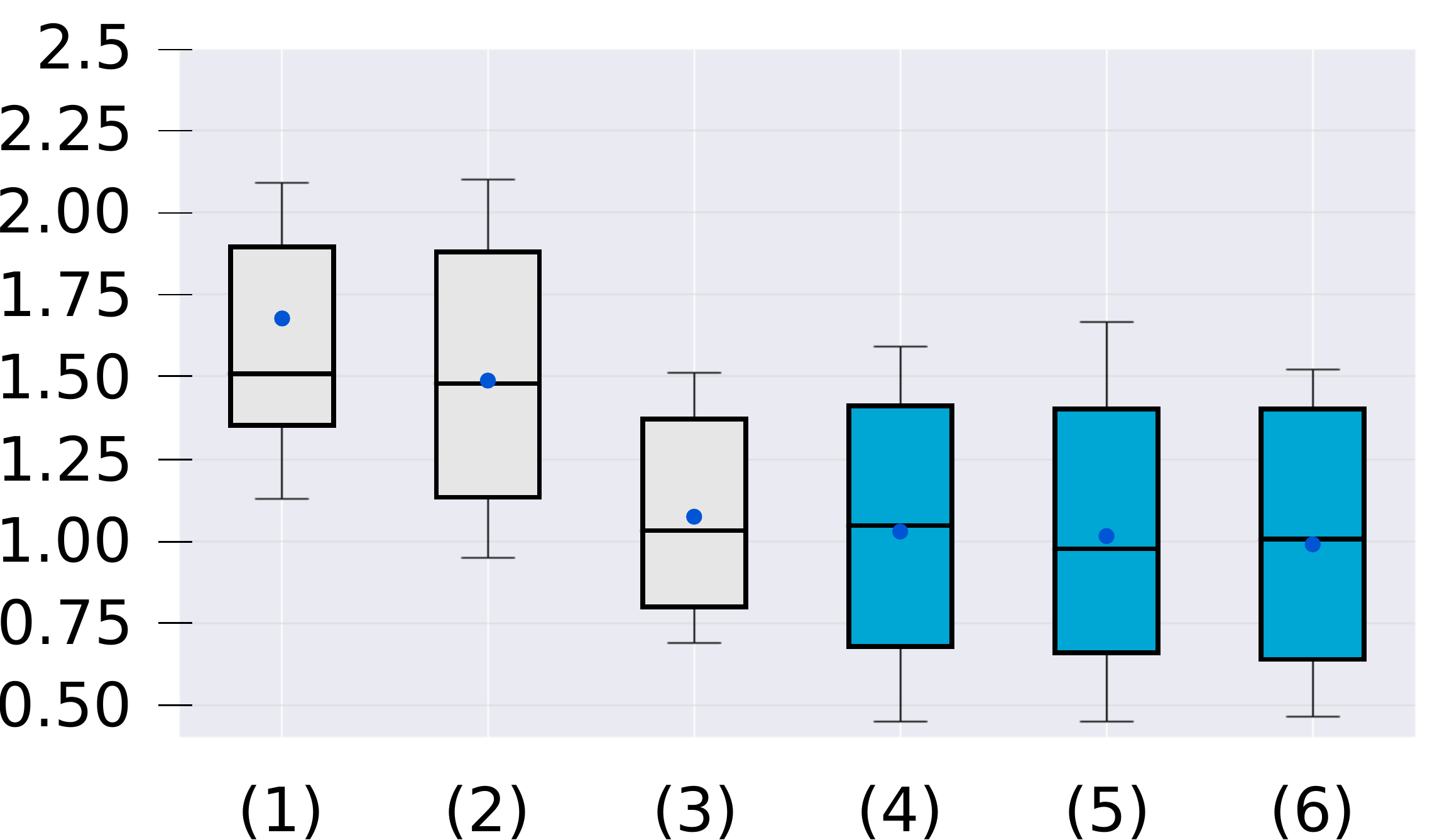}\label{fig:obrpe}}
  \subfigure[BM]{\includegraphics[width=0.47\textwidth]{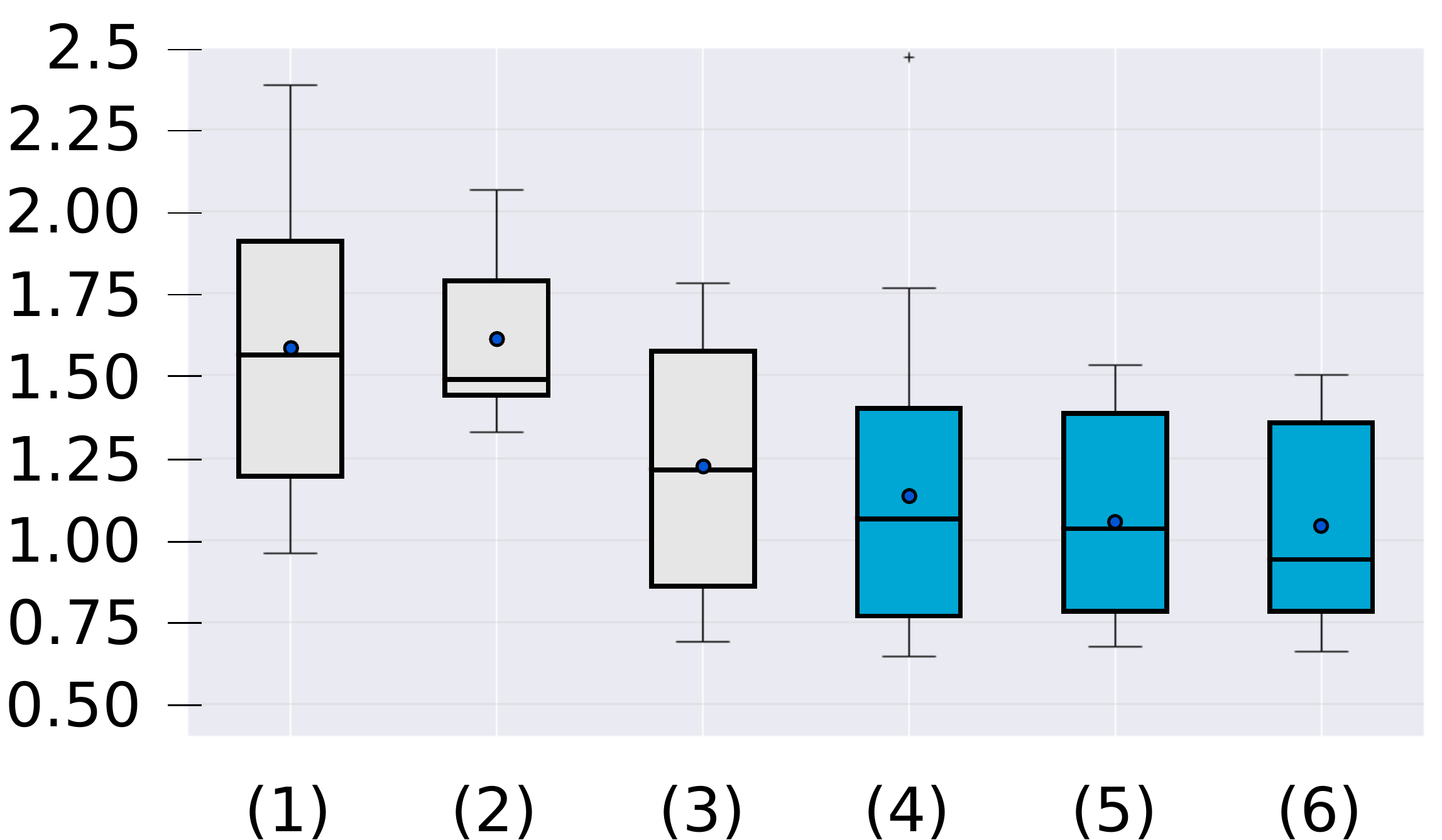}\label{fig:bm}}
\caption{Mean absolute error of OBRPE and BM surface segmentation in pixels on our private data set. (1) Multiclass U-Net with OBRPE/BM targets, (2) Multiclass U-Net with OBRPE/BM/drusen targets, (3) Multitask approach with 2 decoders, (4-6) Our model with (4) disconnected decoders, (5) connected decoders and gradient flow and (6) connected decoders and without gradient flow.}
\label{fig:layer-results}
\end{figure}

\begin{figure}[t!]
    \centering
    \includegraphics[width=0.9\textwidth]{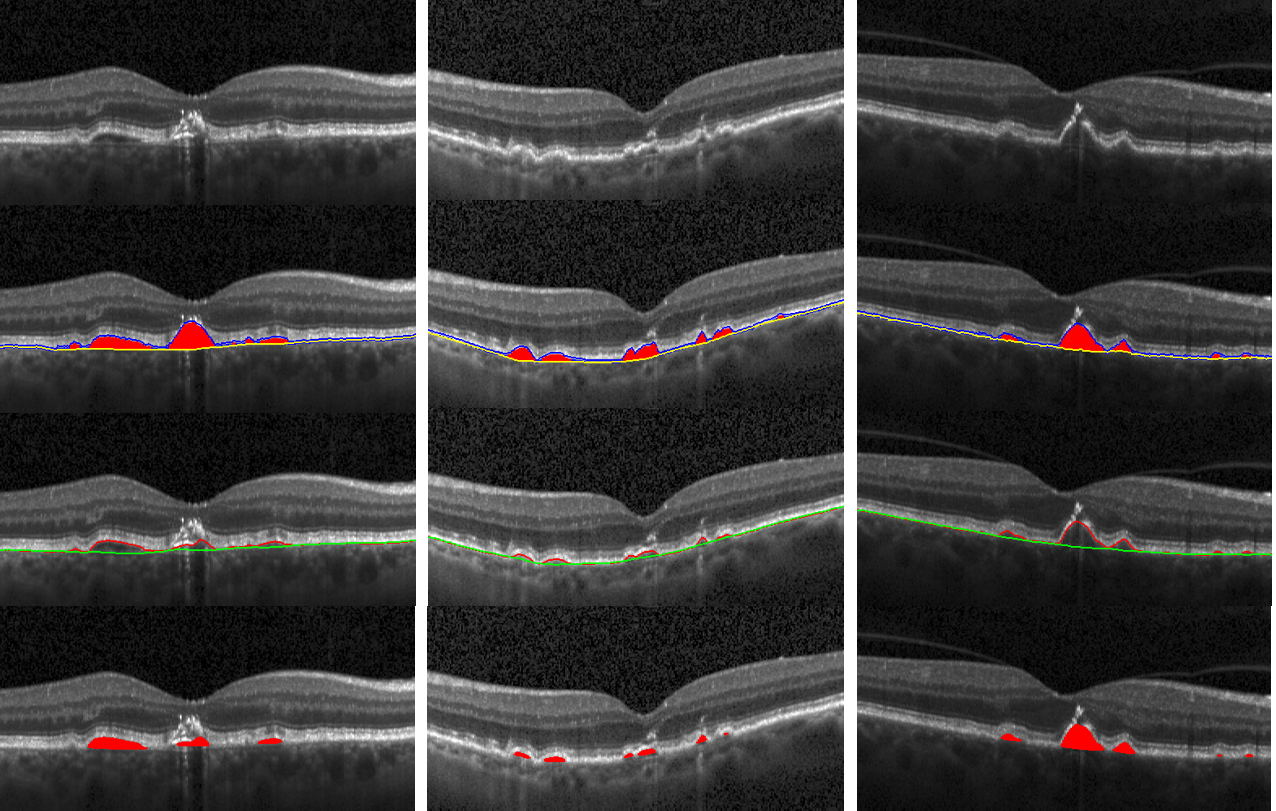}
    \caption{Qualitative results of our best method (connected without gradient flow). From left to right: worst, median and best cases according to volume level Dice (0.4, 0.7, 0.94, respectively). From top to bottom: original B-scan, manual segmentation (blue=OBRPE, red=drusen, yellow=BM), predicted layers (red=OBRPE, green=BM) and predicted drusen (red).}
    \label{fig:qualitative_results}
\end{figure}

Three baselines were quantitatively compared with respect to our three proposed models using our private data set of early/intermediate AMD subjects: a binary U-Net trained for background vs. drusen segmentation~\cite{GorgiZadeh2017}; a multiclass U-Net trained for segmenting OBRPE and BM~\cite{GorgiZadeh2017}; and a multidecoder alternative such as the one in~\cite{playout2018multitask}, with two decoders for segmenting OBRPE and BM, without connections between them. Enough capacity was given to these baselines to match the one of our models. These quantitative results for drusen segmentation are summarized in Table~\ref{table:comparison}. Fig.~\ref{fig:layer-results} depicts boxplots representing the mean absolute error in BM and OBRPE segmentation. Our method performed better than the baselines in any of its forms. Inbound/outbound connections with the drusen decoder were observed to significantly increase performance when no gradient flow is allowed through them (with vs. without gradients, $p < 0.05$). Exemplary results on the central B-scan of volumes with the highest, median and lowest volume level dice are depicted in Fig.~\ref{fig:qualitative_results}. Our approach is able to consistently detect drusen of any size. The median case (second column) presents small material accumulations that are slightly undersegmented by our method. A similar behavior is observed in the worst case (first column) for the large drusen in the center of the image. Nevertheless, both cases present visual ambiguities that are difficult to address even for human readers. 

\begin{table*}[t]  
\centering
\caption{Quantitative evaluation of drusen segmentation results in our private data set. Values are reported across averaged eye-level performance.}
\resizebox{\textwidth}{!}{
\begin{tabular}{C{6cm} | C{2cm} | C{2cm} | C{2cm} }
  \hline
  \textbf{Method} & \textbf{Dice} & \textbf{Precision} &  \textbf{Recall} \\
  \hline
  Binary U-Net~\cite{GorgiZadeh2017} (Drusen) & 0.66$\pm$0.14  & 0.79$\pm$0.09 &  0.58$\pm$0.17  \\
  \hline
  Multiclass U-Net~\cite{GorgiZadeh2017} (OBRPE/BM) & 0.68$\pm$0.13 & 0.79$\pm$0.12 & 0.59$\pm$0.18 \\
  \hline
  Multitask~\cite{playout2018multitask} (OBRPE/BM decoders) & 0.69$\pm$0.18 & 0.79$\pm$0.16 & 0.62$\pm$0.2 \\
  \hline
  Ours (disconnected decoders) & 0.71$\pm$0.13 & 0.83$\pm$0.07 & 0.64$\pm$0.17 \\
  \hline
  Ours (connected with gradient flow) & 0.72$\pm$0.12 & 0.83$\pm$0.08 & 0.65$\pm$0.16 \\
  \hline
  Ours (connected w/o gradient flow) & \textbf{0.73$\pm$0.12} & \textbf{0.84$\pm$0.07} & \textbf{0.67$\pm$0.1} \\
  \hline
\end{tabular}}
\label{table:comparison} 
\end{table*}

Finally, Table~\ref{table:Farsiucontrol} presents an evaluation of our best model in the public data set. To match the available annotations, our approach was trained for ILM, IBRPE and BM segmentation, with one branch per target surface. \H{Our model clearly outperformed two recently proposed methods~\cite{shah2018multiple,liefers2019dense} that reported their performance on the same dataset.}

\begin{table*}[t]  
\centering
\caption{Quantitative evaluation of BM, IBRPE and ILM segmentation in terms of absolute surface differences in pixels on Duke data sets~\cite{farsiu2014quantitative}. This evaluation is done on AMD and control subjects for 100 AMD cases and 100 healthy cases.}
\begin{tabular}{C{5cm} | C{2cm} | C{2cm} | C{2cm} }
  \hline
  \multicolumn{4}{c}{\textbf{AMD set}} \\
  \hline
  \textbf{Method} & \textbf{ILM} & \textbf{IBRPE} &  \textbf{BM}  \\
  \hline
  Shah \textit{et al.}, 2018~\cite{shah2018multiple} &  $1.15\pm0.25$ & $1.88\pm0.57$ &  $1.81\pm0.56$  \\
  \hline
  Liefers \textit{et al.}, 2019~\cite{liefers2019dense} &  $1.055  $ & $1.568  $ &  $1.858  $ \\
  \hline
  Ours (connected w/o gradients) & \textbf{0.88 $\pm$ 0.09} &\textbf{1.23 $\pm$ 0.11} & \textbf{1.15 $\pm$ 0.1} \\
  \hline
  \hline
  \multicolumn{4}{c}{\textbf{Control (healthy) set}} \\
  \hline
  \textbf{Method} & \textbf{ILM} & \textbf{IBRPE} &  \textbf{BM}  \\
  \hline
  Shah \textit{et al.}, 2018~\cite{shah2018multiple} &  $1.04\pm0.07$ & $1.19\pm0.18$ &  $1.54\pm0.31$  \\
  \hline
  Liefers \textit{et al.}, 2019~\cite{liefers2019dense} & $ 0.84 $ & $1.28 $ &  $1.227 $ \\
  \hline
  Ours (connected w/o gradients) & \textbf{0.65 $\pm$ 0.06} &\textbf{1.06 $\pm$ 0.12} & \textbf{0.9 $\pm$ 0.08} \\
  \hline
\end{tabular}
\label{table:Farsiucontrol} 
\end{table*}
\section{Discussion}
\label{sec:discussion}

We introduced a novel multitask approach for multiclass segmentation in retinal OCT images. In particular, we showed that our multi-decoder architecture is able to outperform standard baselines by incorporating an intermediate decoder that targets the area between two stacked interfaces. This improvement, for example, is observed in drusen segmentation, with an increment of 3\% in Dice index with respect to the multitask approach presented in~\cite{playout2018multitask}. Introducing connections between decoders also allowed for further improvement on Dice for drusen segmentation, while also improving performance for OBRPE and BM segmentation. Surprisingly, the best results were observed when no gradient flow was allowed between the connected decoders. This implies that, at least on our data set, local information provided by neighbouring classes can improve results. Finally, we also include an evaluation for layer segmentation in OCT images using the Duke public available data set, to further compare our method with the state-of-the-art. We used the same portion size of data as this work, clearly reported the lowest error for all the evaluated surfaces. This promising empirical evidence leads us to envision further applications that might benefit from our approach, such as layer segmentation in retinal diseased cases with fluid, where an extra decoder can be added targeting only those lesions.


\noindent\\
\textbf{Acknowledgements}\\
This work was funded by the Christian Doppler Research Association, the Austrian Federal Ministry for Digital and Economic Affairs and the National Foundation for Research, Technology and Development. We thank the NVIDIA corporation for a GPU donation. JIO is funded by WWTF AugUniWien/FA7464A0249 (Medical University of Vienna); VRG12-009 (University of Vienna).

%
%
%
%
\bibliographystyle{splncs}
\bibliography{main}

\end{document}